# Nodeless superconducting gap and electron-boson coupling in $(La,Pr,Sm)_3Ni_2O_7$ films

Jianchang Shen[1†], Guangdi Zhou[2†], Yu Miao[1,3†], Peng Li[2,4†], Zhipeng Ou[1,3,5†], Yaqi Chen[2], Zechao Wang[2], Runqing Luan[1,3,5], Hongxu Sun[1,3,5], Zikun Feng[1,3,5], Xinru Yong[1,3,5], Yueying Li[2], Lizhi Xu[2], Wei Lv[2], Zihao Nie[2], Heng Wang[2,4], Haoliang Huang[2,4], Yu-Jie Sun[2,4], Qi-Kun Xue[2,4,6]*, Junfeng He[1,3,5]*, Zhuoyu Chen[2,4]*

[1]Department of Physics and CAS Key Laboratory of Strongly-coupled Quantum Matter Physics, University of Science and Technology of China, Hefei, Anhui 230026, China

[2]State Key Laboratory of Quantum Functional Materials, Department of Physics, and Guangdong Basic Research Center of Excellence for Quantum Science, Southern University of Science and Technology, Shenzhen 518055, China

[3]Hefei National Laboratory, University of Science and Technology of China, Hefei 230088, China

[4]Quantum Science Center of Guangdong-Hong Kong-Macao Greater Bay Area, Shenzhen 518045, China

[5]Hefei National Research Center for Physical Sciences at the Microscale, University of Science and Technology of China, Hefei, 230026, China

[6]Department of Physics, Tsinghua University, Beijing 100084, China

†These authors contributed equally.

*Corresponding author.

Email: chenzhuoyu@sustech.edu.cn, jfhe@ustc.edu.cn, xueqk@sustech.edu.cn

The discovery of superconductivity in Ruddlesden-Popper (RP) bilayer nickelate films under ambient pressure provides an unprecedented opportunity to directly investigate electronic energy scales of the superconducting state and the pairing mechanism. Here, we report angle-resolved photoemission spectroscopy measurements of superconducting $(La,Pr,Sm)_3Ni_2O_7$ thin films epitaxially grown on $SrLaAlO_4$ substrates by developing an ultra-high vacuum low-temperature quenching and transfer technique. A finite superconducting gap of ~18 meV with pronounced coherence peak is observed along the Brillouin zone diagonal direction. Remarkably, the finite superconducting gap persists across the entire Brillouin zone of the underlying Fermi surfaces, revealing the absence of gap nodes. An abrupt band renormalization, manifested as a kink in the energy-momentum dispersion at ~70 meV below the Fermi level, indicates an electron-boson coupling in the system. The simultaneous observation of a nodeless superconducting gap and electron-boson coupling provides crucial insights into the pairing symmetry and gluing mechanism in high-$T_C$ RP bilayer nickelates.

The discovery of superconductivity in nickelates opens a new frontier in condensed matter physics [1-28]. Among the recently reported nickelate superconductors, the square-planar nickelates share similar ~3$d^9$ configuration with cuprate high-temperature superconductors [1-3,7,8], whereas the Ruddlesden-Popper (RP) nickelates hold a nominal ~3$d^{7.5}$ configuration [4-6]. High-pressure measurements on the bilayer nickelates indicate a superconducting transition temperature ($T_C$) at around the liquid nitrogen temperature [4,10-13], raising a critical issue on the superconducting mechanism.

As established in both cuprates and iron-based high-$T_C$ superconductors, experimental observations of electronic energy scales of the superconducting state are crucial for understanding of the physical mechanism. In particular, the superconducting gap reveals the energy to break electron Cooper pairs [29-31], and the electron-boson coupling, manifested as a dispersion kink, may be linked to the superconducting pairing glue [30-32]. Recently, superconductivity has been reported in RP phase bilayer nickelate thin films under ambient pressure [17,18], providing an opportunity to directly probe these energy scales from the electronic structure and study the pairing symmetry by angle-resolved photoemission spectroscopy (ARPES).

High-resolution ARPES measurements on superconducting bilayer nickelate films are challenging. Warming the superconducting films above 200 K in vacuum would inevitably induce oxygen loss and thus destroy their superconductivity. To overcome this problem, we have developed an ultra-high vacuum (UHV) low-temperature quenching and transfer technique. The samples are immediately quenched to a temperature at ~120 K after growth. The low temperature and thus the superconductivity of the films are maintained during the sample transfer and ARPES measurements. As a result, we have successfully carried out long-distance (over 1200 km) sample transfer and performed measurements on the superconducting $(La,Pr,Sm)_3Ni_2O_7$ films grown in Shenzhen by using high-resolution laser-based ARPES system in Hefei and synchrotron-based ARPES system in Shanghai. We observe two striking results, namely a nodeless superconducting gap implying s-wave (including s±) pairing symmetry, and a dispersion kink indicating electron-boson coupling.

Superconducting $(La,Pr,Sm)_3Ni_2O_7$ films grown on $SrLaAlO_4$ substrates are examined by scanning transmission electron microscopy (STEM) over a large field of view (Fig. 1A), demonstrating the homogeneity and pure phase of the samples (see also Fig. S1 for reflective high-energy electron diffraction and x-ray diffraction characterizations). The $SrTiO_3$ capping layer is only deposited on the STEM samples for protection, but not on the samples for ARPES and transport experiments. The resistance measurement exhibits a superconducting onset temperature $T_C^{onset}$ ~ 46 K and a zero resistance temperature $T_C^{zero}$ ~16 K (Fig. 1B). As shown in Fig. 1 (C-H), laser-based ARPES measurements reveal a well-defined superconducting gap. This is first evidenced by the Fermi-Dirac divided energy distribution curves (EDCs) near the Fermi momentum $k_F$ (Figs. 1D, 1E), which exhibit a back-bending behavior with the band-top below the Fermi level ($E_F$) (Fig. 1E). Superconducting coherent peaks are resolved in these EDCs (Fig. 1E), enabling a quantitative determination of the gap magnitude. As shown in Fig. 1F for a Brillouin zone (BZ) diagonal direction, the symmetrized EDC at $k_F$ with clear coherent peaks can be fitted to a phenomenological model typically used in cuprate superconductors [33]. The fitting yields a gap size of ~18 meV. Importantly, the superconducting coherent peak and finite superconducting gap are observed in momentum regions at and offset from the BZ diagonal (Figs. 1G, 1H), indicating a nodeless gap structure. This is in sharp contrast to many cuprates [42], where a gap node is always observed along the diagonal (a key feature of *d*-wave gap symmetry). This nodeless feature is confirmed in multiple superconducting film samples with different thicknesses and chemical contents.

Due to the relative low photon energy (7 eV), laser-based ARPES measurements are limited to the region near the BZ diagonal. To probe the energy gap near the BZ boundary, synchrotron-based ARPES measurements are carried out (Figs. 2A-C). For clarity, we follow the notation convention in the earlier studies [34,35] and label the underlying Fermi surfaces and their associated bands as $\alpha$, $\beta$ and $\gamma$, respectively. While the expected $\alpha$ band is strongly suppressed in the first BZ due to matrix element effect [34], both $\beta$ and $\gamma$ bands are observed (Figs. 2A, 2B). An energy gap is identified on both $\beta$ and $\gamma$ bands, evidenced by the EDCs at Fermi momenta (Fig. 2C and supplementary Fig. S2). We summarize the momentum evolution of the gap on both bands in Figs. 2D-E, and visualize the results in Fig. 2F. It is clear that a gap node is absent along both the zone diagonal and boundary, inconsistent with the $d_{x2-y2}$ and $d_{xy}$ gap symmetry scenarios, respectively. The gap magnitude shows moderate momentum dependent variations along the $\beta$ Fermi surface, but exhibits a more isotropic behavior along the $\gamma$ Fermi surface within our experimental error bars.

Temperature-dependent measurement of the superconducting gap is performed along the (0,0)-(π,π) direction (Fig. 3 and supplementary Figs. S3-S4). We quantify spectral weight suppression near $E_F$, following a typical analysis method used in cuprates [36,37]. The analysis proceeds in three steps: First, we obtain momentum-integrated EDCs (Fig. 3A demonstrates a 150-K case, also see supplementary Fig. S4). Second, we subtract the fitted line of 150-K EDC from EDCs at different temperatures to isolate the suppression feature (Fig. 3B). Third, the quantified spectral weight suppression $A(E_F)$ is calculated by integrating the difference curve within a small window (-2 meV, 0 meV) near $E_F$ (illustrated in Fig. 3B for the 10 K data). The resulting temperature evolution of $A(E_F)$ quantitatively tracks the suppression of electron density of states (DOS) associated with the energy gap (Fig. 3C). $A(E_F)$ at a temperature $T$ (labeled as $A_T$) is normalized by $A(E_F)$ at 10 K, using the equation $A_T/|A_{10K}|$. A sudden drop of electron DOS is clearly identified at $T_C^{onset}$, consistent with previous observations in cuprates [36,37]. This sharp decrease of DOS at $T_C^{onset}$ directly links this feature to superconductivity, and confirms that the energy gap with coherence peaks observed at low temperatures is indeed of superconducting origin. In the meantime, we note that a more gradual suppression of the DOS persists to an elevated temperature above $T_C^{onset}$ (Fig. 3C and supplementary Fig. S3), indicating the existence of a pseudogap. This result implies that preformed Cooper pairs and possible competing orders are potentially important in this system.

After revealing the energy gap, we investigate another energy scale in the electron band dispersion. Laser-based photoemission measurement along (0,0)-(π,π) diagonal direction of the BZ at 10 K (Fig. 4A) uncovers a change in the dispersion slope below the $E_F$ (Fig. 4A). This feature becomes more evident when the dispersion is extracted from the raw momentum distribution curves (MDCs) and plotted in an expanded scale in Fig. 4B (also see supplementary Fig. S5). A dispersion kink is discernable at ~70 meV, accompanied by distinct electron velocities below and above the kink energy. This is unexpected in the first-principles calculations [34], in which a nearly constant electron velocity is expected in this energy region (Fig. 4E). Effective real part of electron self-energy (Re$\Sigma$) is extracted from the dispersion by taking a straight line as an effective bare band [30,31]. A peak is identified at the same energy scale regardless of the selection of the effective bare band (Fig. 4C). At this energy, the MDC width shows an abrupt change (Fig. 4D), and the raw EDCs also display an obvious step feature (Fig. 4F). We note that this dispersion kink is universal in all superconducting films we measured (e.g. see supplementary Fig. S6 for the results on another sample).

Below we discuss possible implications of our observations. Firstly, a finite superconducting gap is observed along both the BZ diagonal and boundary, demonstrating the absence of gap nodes in the superconducting state of the RP nickelate bilayers, distinct from that in cuprates [30,31,40-42]. It would be interesting to examine whether it is relevant to the absence of charge transfer layer in the RP bilayer nickelates. Secondly, the quantitative determination of the superconducting gap enables an estimation of the gap to $T_C$ ratio. The extracted ratio $2\Delta/k_BT_C \sim 9$ is much greater than the conventional value. Thirdly, the observed dispersion kink reveals an energy scale at ~70 meV below $E_F$. A similar phenomenon has been widely reported in cuprates and interpreted as electron-boson coupling [30,31]. While it is believed that the electron-boson coupling is intimately associated with high $T_C$ superconductivity, the nature of the bosonic mode and the role it plays in mediating/breaking Cooper pairs have been central debates in cuprates [30,31,38,39]. Our observations establish that electron-boson coupling is a significant factor in bilayer nickelate superconductors. Further experiments are required to determine whether this coupling plays a dominant role in the pairing mechanism. For instance, future studies could quantitatively examine the correlation between the electron-boson coupling strength and the magnitude of the superconducting gap. To specifically test for an electron-phonon contribution, isotope effect measurements are encouraged. Finally, we would like to point out that the nodeless gap and electron-boson coupling observed in the current nickelate superconducting films were also reported in monolayer FeSe superconducting films on $SrTiO_3$ substrates [43-46]. Shared features in these two systems could provide critical insights for understanding superconducting pairing mechanism in different high $T_C$ superconductors.

In summary, by developing an UHV low-temperature quenching and transfer technique, we have successfully carried out ARPES measurements on superconducting Ruddlesden-Popper (RP) bilayer nickelate films. The observed superconducting gap, dispersion kink and pseudogap in superconducting $(La,Pr,Sm)_3Ni_2O_7$ films provide crucial information on the critical energy scales and thus the pairing mechanism of high temperature superconductivity in RP bilayer nickelates.

References：

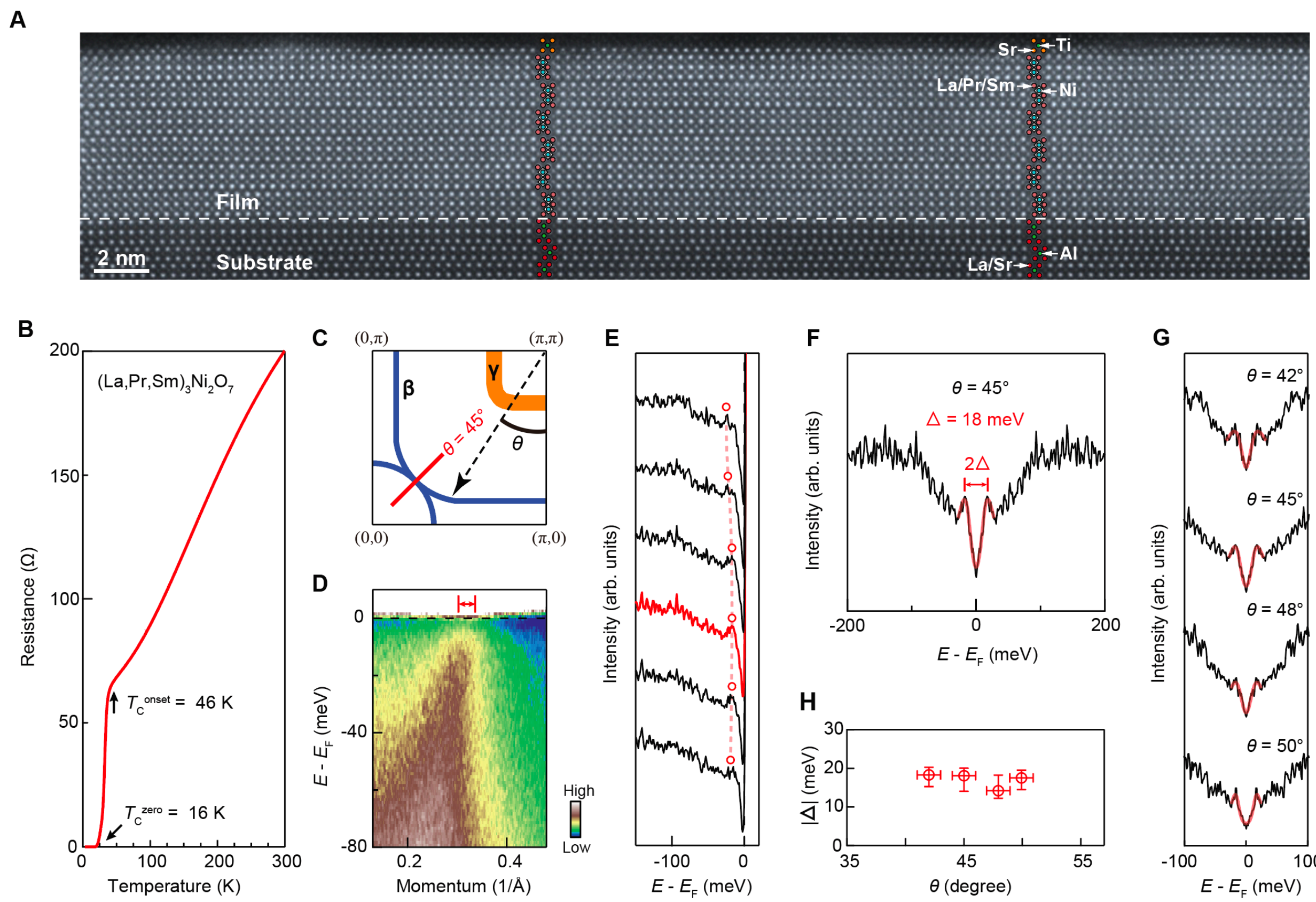

**Fig. 1. Superconducting gap near the Brillouin zone diagonal.** (**A**) STEM high angle annular dark field (HAADF) image over a large field of view. The $SrTiO_3$ cap is only grown on STEM samples for protection, but not on samples for ARPES and transport. (**B**) Resistance measurement of a 3-unit-cell (3UC) thick $La_{2.31}Pr_{0.24}Sm_{0.45}Ni_2O_7/SrLaAlO_4$ thin film. The arrows indicate the superconducting onset temperature and zero resistance temperature, respectively. The current channel is 5 mm wide. The distance between two voltage measurement electrodes is 0.7 mm. (**C**) Schematic of the underlying Fermi surface. (**D**) Photoelectron intensity plot of the band along (0,0)-(π,π) direction of a superconducting 3UC $La_{2.31}Pr_{0.24}Sm_{0.45}Ni_2O_7/SrLaAlO_4$ thin film, shown in an expanded scaled near $E_F$. The Fermi-Dirac function is removed. (**E**) Fermi-Dirac divided EDCs in a region near the Fermi momentum $k_F$, marked by the red arrows in (D). The EDC at $k_F$ is shown in red. (**F**) Symmetrized EDC at $k_F$. The red curve is the fitting result by a phenomenological model. (**G**) Symmetrized EDCs at Fermi momenta along the underlying $\beta$ Fermi surface, the momentum locations of the EDCs are defined by the angle $\theta$ in (C). The fitting results are shown in red. (**H**) The extracted gap size as a function of angle $\theta$. The error bars represent the uncertainties in the extraction of the gap. The measurements are performed at 10 K.

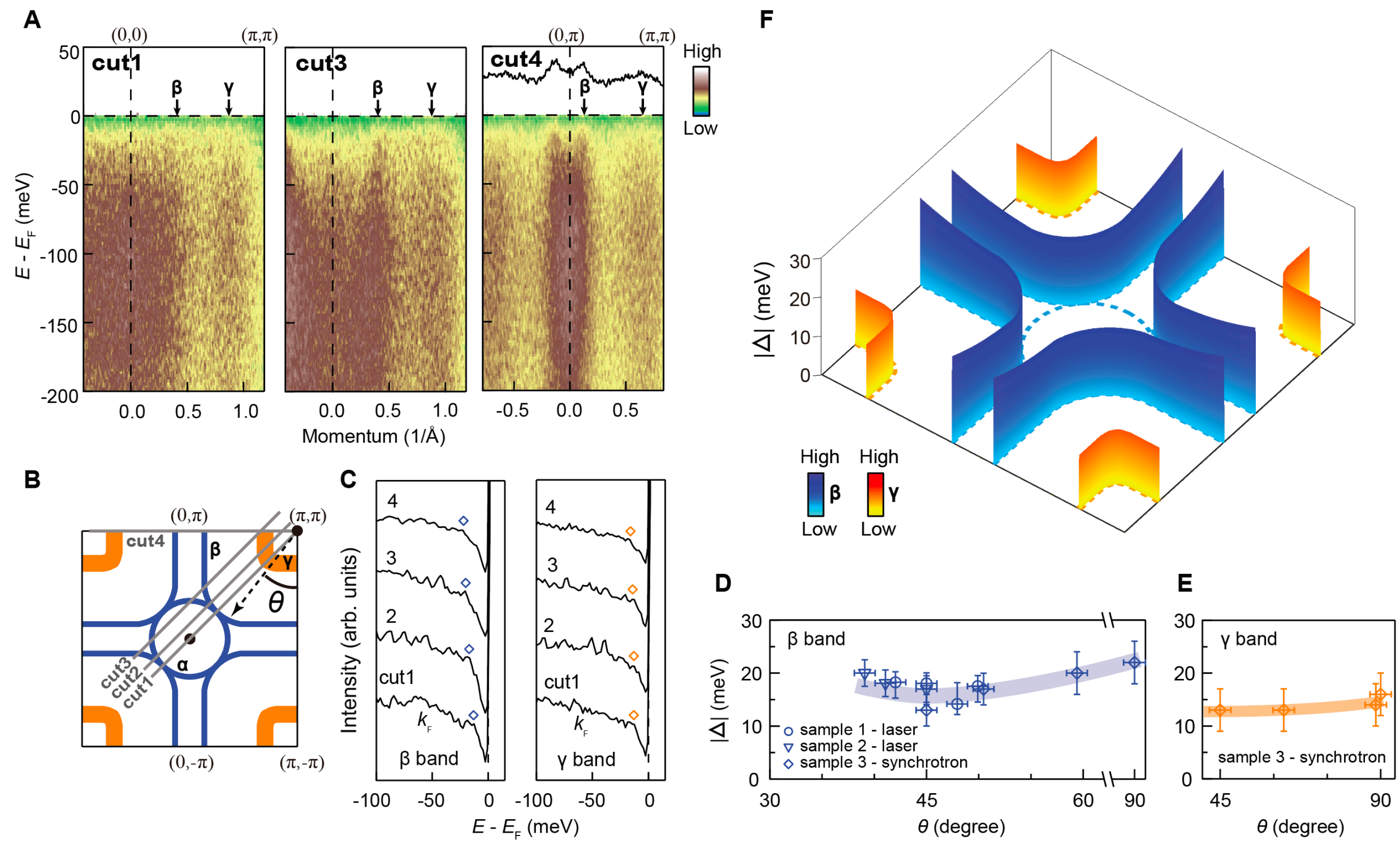

**Fig. 2. Momentum dependence of the superconducting gap.** (**A**) Photoelectron intensity plot along three momentum cuts of a superconducting 2UC $La_{2.85}Pr_{0.15}Ni_2O_7/SrLaAlO_4$ thin film, probed by 63 eV synchrotron light. (**B**) Schematic of the underlying Fermi surface and locations of the momentum cuts. (**C**) Fermi-Dirac divided EDCs at the Fermi momenta ($k_F$) crossed by cuts 1-4 on the $\beta$ and $\gamma$ Fermi surface, respectively. (**D-E**) Extracted magnitude of the superconducting gap as a function of angle $\theta$ for three samples (circles: 3UC $La_{2.31}Pr_{0.24}Sm_{0.45}Ni_2O_7/SrLaAlO_4$, laser-ARPES same as Fig.1; triangles: 2UC $La_{2.31}Pr_{0.24}Sm_{0.45}Ni_2O_7/SrLaAlO_4$ laser-ARPES; diamonds: 2UC $La_{2.85}Pr_{0.15}Ni_2O_7/SrLaAlO_4$ synchrotron-ARPES). Since La, Pr, and Sm all manifest the same +3 valence state in these compounds, these three sample configurations exhibit similar $T_C^{onset}$ above 40 K. The definition of $\theta$ is shown in (B). The blue and orange shades are a guide to the eye. The error bars represent the uncertainties in the extraction of the energy gap. (**F**) Visualization of the momentum dependent energy gap.

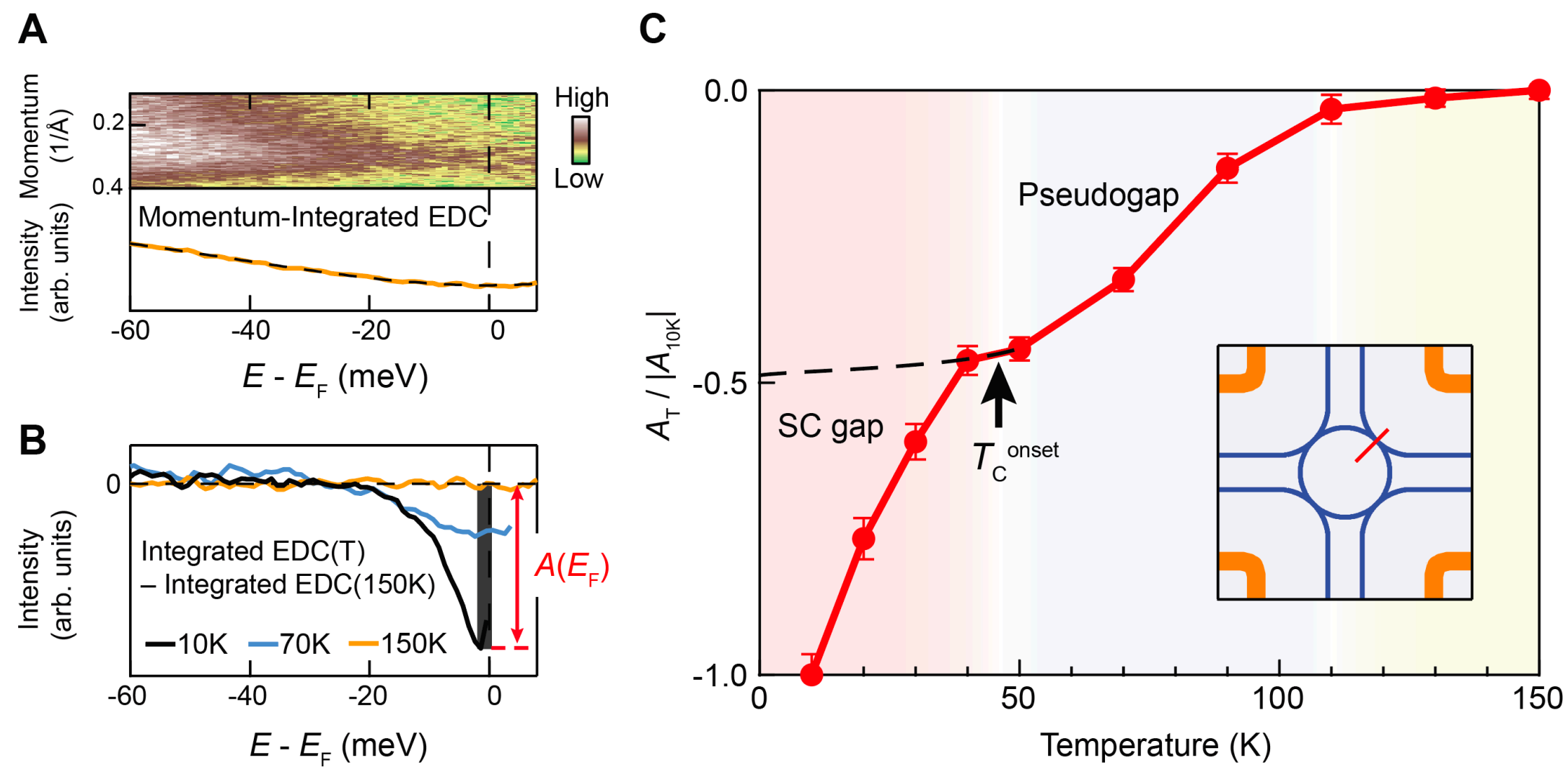

**Fig. 3. Temperature evolution of the spectral weight.** (**A**) Photoelectron intensity plot of the band structure along the (0,0)-(π,π) direction, measured on a superconducting 3UC $La_{2.31}Pr_{0.24}Sm_{0.45}Ni_2O_7/SrLaAlO_4$ thin film at 150 K (top). The momentum location of the cut is marked by the red line in the inset of (C). Momentum-integrated EDC (bottom). The dashed line represents a polynomial fitting to the integrated EDC. (**B**) Representative momentum-integrated EDCs at 10 K, 70 K and 150 K, subtracted by the fitting curve of the momentum-integrated EDC at 150 K. All EDCs are normalized between -0.2 eV and -0.15 eV. The momentum-integrated intensity at zero energy [$A(E_F)$] is defined by the area of the dark gray region. The energy window is between -2 meV and $E_F$. (**C**) Temperature evolution of $A(E_F)$. For comparison, $A(E_F)$ at a temperature $T$ (labeled as $A_T$) is normalized by the absolute value of $A(E_F)$ at 10 K (labeled as $|A_{10K}|$).

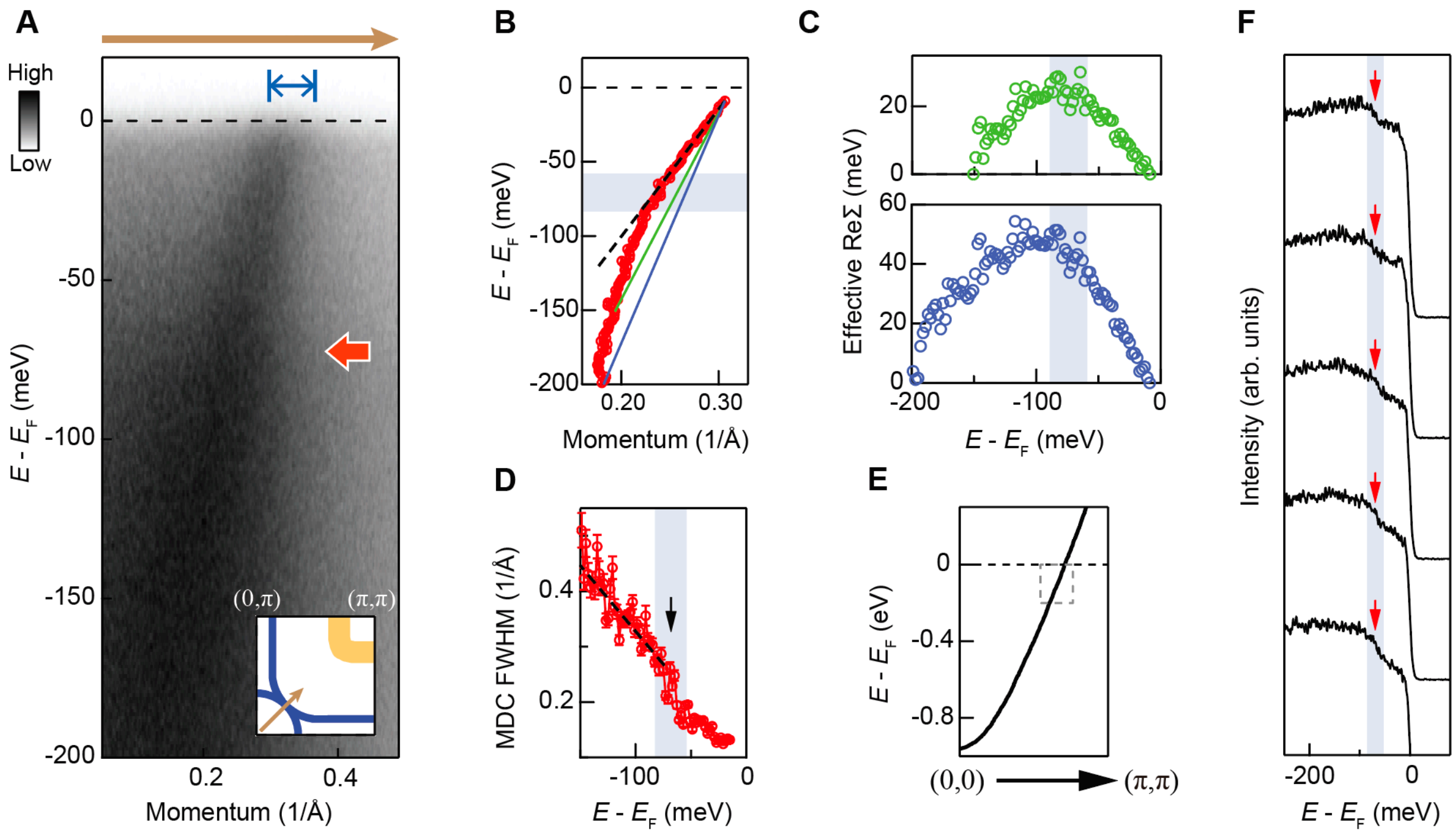

**Fig. 4. Electron energy band along the (0,0)-(π,π) direction measured at 10 K.** (**A**) Photoelectron intensity plot as a function of energy and momentum measured at 10 K along the (0,0)-(π,π) direction, marked by the brown arrow in the inset. The red arrow indicates the kink position. (**B**) MDC-derived dispersion from (A). The dashed line is a guide for the eye. The green and blue lines represent two effective bare bands. (**C**) Effective real part of electron self-energy, obtained by subtracting the MDC-derived dispersion from the two effective bare bands, respectively. (**D**) Full-width-at-half-maximum (FWHM) of the MDC as a function of energy. (**E**) Calculated band structure of the $d_{x2-y2}$ band along the (0,0)-(π,π) direction (from ref. 34). The dashed box indicates the energy region studied in (A) and (B). (**F**) Raw EDCs in a region near the Fermi momentum $k_F$, marked by the blue arrows in (A). The red arrows indicate the step feature in the EDCs. The light blue region in (B-D and F) marks the energy feature at ~70 meV.